\documentclass[a4paper,11pt]{article}
\usepackage{pos}
\usepackage{setspace}

\usepackage{cleveref}
\Crefformat{equation}{Eq.~(#2#1#3)}
\Crefmultiformat{equation}{Eqs.~(#2#1#3)}{ and~(#2#1#3)}{, (#2#1#3)}{ and~(#2#1#3)}
\Crefformat{figure}{Fig.~#2#1#3}
\Crefformat{section}{Sec.~#2#1#3}

\newcommand{\bv}[1]{{\mathbf{#1}}}

\def\llangle{\left\langle}
\def\rrangle{\right\rangle}
\newcommand{\ea}[1]{{\llangle #1 \rrangle}}

\title{Investigating the low moments of the nucleon structure functions in lattice QCD}

\author*[a,1]{K.U. Can}
\author[a]{A.~Hannaford-Gunn}
\author[a]{E.~Sankey}
\author[b]{R.~Horsley}
\author[c]{Y.~Nakamura}
\author[d]{H.~Perlt}
\author[e]{P.E.L.~Rakow}
\author[f]{G.~Schierholz}
\author[g]{H.~St\"{u}ben}
\author[a]{R.D.~Young}
\author[a]{J.M.~Zanotti}

\affiliation[a]{CSSM, Department of Physics, The University of Adelaide,
Adelaide SA 5005, Australia.}
\affiliation[b]{School of Physics and Astronomy, University of Edinburgh,
Edinburgh EH9 3JZ, UK.}
\affiliation[c]{RIKEN Center for Computational Science,
Kobe, Hyogo 650-0047, Japan.}
\affiliation[d]{Institut f\"{u}r Theoretische Physik, Universit\"{a}t Leipzig,
04103 Leipzig, Germany.}
\affiliation[e]{Theoretical Physics Division, Department of Mathematical Sciences, University of Liverpool, Liverpool L69 3BX, United Kingdom.}
\affiliation[f]{Deutsches Elektronen-Synchrotron DESY, 
Notkestr. 85, 22607 Hamburg, Germany.}
\affiliation[f]{Regionales Rechenzentrum, Universit\"{a}t Hamburg,
20146 Hamburg, Germany.}

\note{For the CSSM-QCDSF/UKQCD Collaborations}

\emailAdd{kadirutku.can@adelaide.edu.au}

\abstract{We highlight QCDSF/UKQCD Collaboration's recent developments on computing the Compton amplitude directly via an implementation of the second order Feynman-Hellmann theorem. As an application, we compute the nucleon Compton tensor across a range of photon virtuality at an unphysical quark mass. This enables us to study the $Q^2$ dependence of the low moments of the nucleon structure functions in a lattice calculation for the first time. We present some selected results for the moments of the $F_1$, $F_2$ and $F_L$ structure functions and discuss their implications.}

\FullConference{%
 The 38th International Symposium on Lattice Field Theory, LATTICE2021
  26th-30th July, 2021
  Zoom/Gather@Massachusetts Institute of Technology
}

\begin{document}
\maketitle

\section{Introduction}
Computing nuclear structure functions poses several challenges to lattice QCD practitioners, most notably the operator mixing and renormalisation issues~\cite{Martinelli:1996pk,Beneke:1998ui} that complicate the lattice operator product expansion (OPE) approach. 

Recently, the focus has largely been directed to light-cone PDFs that can be computed from quasi- and pseudo-PDF approaches which evade these issues and obtain the $x$-dependence of the parton distributions. A detailed account of quasi- and pseudo-PDF, and other approaches, including their limitations, and in general what has been accomplished so far, is given in recent reviews~\cite{Lin:2017snn,Cichy:2018mum} and plenaries at the lattice conferences~\cite{Constantinou:2020pek,Cichy:2021lat} highlighting the immense efforts and the progress of the lattice community.

A complementary approach that is pursued by the QCDSF/UKQCD Collaboration is to extract the forward Compton amplitude and access the structure functions that way. This approach is akin to an experimental one where we are able to obtain the full Compton amplitude, which includes all twist contributions and power corrections. Our determination of the Compton amplitude takes advantage of the Feynman-Hellmann approach to hadron structure~\cite{Chambers:2017tuf,Bouchard:2016heu}. The procedure is detailed in \cite{PhysRevLett.118.242001,PhysRevD.102.114505} which presents a derivation for the determination of the forward Compton amplitude via the Feynman-Hellmann theorem, together with an application to the nucleon structure function $F_1$. In this contribution, we summarise the method and report on our recent progress on accessing the $F_1$, $F_2$ and $F_L$ structure functions of the nucleon.  

\section{Compton tensor and the structure functions}
The starting point is the forward Compton amplitude described by the time ordered product of electromagnetic currents sandwiched between nucleon states, 
\begin{align} 
    \label{eq:compamp}
    T_{\mu\nu}(p,q) =& \int d^4z\, e^{i q \cdot z} \rho_{s s^\prime} \ea{p,s^\prime \left| 
    \mathcal{T}\left\{ \mathcal{J}_\mu(z) \mathcal{J}_\nu(0) \right\} \right|p,s},
\end{align}
where $p$ ($s$) is the momentum (spin) of the nucleon, $q$ is the momentum of the virtual photon, and $\rho$ is the polarisation density matrix. We are interested in the unpolarised part of the Compton tensor, which is parametrised in terms of two Lorentz-invariant scalar functions, $\mathcal{F}_1$ and $\mathcal{F}_2$ as follows
\begin{align} 
    \label{eq:compamp_tensor}
    T_{\mu\nu}(p,q) =& \left( -g_{\mu\nu} + \frac{q_\mu q_\nu}{q^2} \right) \mathcal{F}_1(\omega,Q^2) + \left( p_\mu - \frac{p \cdot q}{q^2}q_\mu \right) \left( p_\nu - \frac{p \cdot q}{q^2}q_\nu \right) \frac{\mathcal{F}_2(\omega,Q^2)}{p \cdot q},
\end{align}
where $Q^2 = -q^2$. These invariant Compton structure functions $\mathcal{F}_{1,2}$ are related to the corresponding ordinary structure functions via the optical theorem, which states
$\operatorname{Im}\mathcal{F}_{1,2}(\omega,Q^2) = 2\pi F_{1,2}(x,Q^2)$.
Making use of analyticity, crossing symmetry and the optical theorem, we can write a dispersion relation for $\mathcal{F}$ and connect them to the inelastic structure functions,
\begin{equation} \label{eq:compomega12}
    \overline{\mathcal{F}}_1(\omega,Q^2)= 2\omega^2 \int_0^1 dx \frac{2x \, F_1(x,Q^2)}{1-x^2\omega^2-i\epsilon}, \quad
    \mathcal{F}_2(\omega,Q^2)= 4\omega \int_{0}^1 dx\, \frac{F_2(x,Q^2)}{1-x^2\omega^2-i\epsilon},
\end{equation}
where we will use $\overline{\mathcal{F}}_i(\omega,Q^2) = \mathcal{F}_i(\omega,Q^2)-\mathcal{F}_i(0,Q^2)$ throughout to denote a once subtracted function. Additionally, a once-subtracted dispersion relation for the longitudinal structure function $F_L(x)$ is written as,
\begin{align}\label{eq:compomegaL}
    \overline{\mathcal{F}}_L(\omega,Q^2) &= 2\omega^2 \int_0^1 dx \frac{F_L(x,Q^2)}{1 - x^2 \omega^2 -i\epsilon}, \; \text{where} \\
    \label{eq:FL_x}
    F_L(x,Q^2) &= \left( 1 + \frac{4 M^2}{Q^2} x^2 \right) F_2(x,Q^2) - 2xF_1(x,Q^2),
\end{align}
with $M$ the mass of the nucleon. Note that a subtraction is necessary, given the high-energy behaviour of $F_1$. Although we are only concerned with subtracting it away, understanding the subtraction function is an interesting subject in itself. Related discussions on the subtraction function can be found in~\cite{Walker-Loud:2012ift,Hagelstein:2020awq,Lozano:2020qcg,HannafordSankey:2021lat}, and further details and investigations will be presented in future publications~\cite{HannafordSankey:2021ea,Can:2021fotl}.    

The expression in \Cref{eq:FL_x} recovers the well-known Callan-Gross relation, $F_L(x) = F_2(x) - 2xF_1(x)$, in the $Q^2 \to \infty$ limit. However the $1/Q^2$ term, which picks up the next moment of $F_2$, is crucial for the low- and mid-$Q^2$ regions. With our chosen parametrisation (\Cref{eq:compamp_tensor}), the longitudinal structure function $\mathcal{F}_L$ can be constructed via the following combination of $\mathcal{F}_1$ and $\mathcal{F}_2$,
\begin{equation} \label{eq:FL_comp}
    \mathcal{F}_L(\omega, Q^2) = -\mathcal{F}_1(\omega,Q^2) + \frac{\omega}{2} \mathcal{F}_2(\omega,Q^2) + \frac{2 M^2}{Q^2} \frac{\mathcal{F}_2(\omega,Q^2)}{\omega}.
\end{equation}

With a judicious choice of kinematics, we are able to isolate the Compton structure functions from the tensor in \Cref{eq:compamp}. Working in Minkowski space with metric $g_{\mu\nu} = \operatorname{diag}(+,-,-,-)$, we have
\begin{align}
    \label{eq:compF1}
    \mathcal{F}_1(\omega, Q^2) &= T_{33}(p,q), &&\text{for} \, \mu=\nu=3 \, \text{and} \, p_3=q_3=0, \\
    \label{eq:compF2}
    \mathcal{F}_2(\omega,Q^2) &= \frac{\omega Q^2}{2 E_N^2} \left[ T_{00}(p,q) + T_{33}(p,q) \right],
    && \text{for} \, \mu=\nu=0 \, \text{and} \, p_3=q_3=q_0=0.
\end{align}

Writing \Cref{eq:compomega12,eq:compomegaL} at fixed $Q^2$ as a geometric series, the Compton structure functions can be expanded as an infinite sum of Mellin moments of the inelastic structure functions,
\begin{align} 
    \label{eq:ope_moments1}
    \overline{\mathcal{F}}_1(\omega,Q^2)&=\sum_{n=1}^\infty 2\omega^{2n} M^{(1)}_{2n}(Q^2), &&\text{with} \; M^{(1)}_{2n}(Q^2)= 2\int_0^1 dx\, x^{2n-1} F_1(x,Q^2), \\
    \label{eq:ope_moments2}
    \mathcal{F}_2(\omega,Q^2)&= \sum_{n=1}^\infty 4\omega^{2n-1} M^{(2)}_{2n}(Q^2), &&\text{with} \; M^{(2)}_{2n}(Q^2)= \int_{0}^1 dx\,x^{2n-2} F_2(x,Q^2), \\
    \label{eq:ope_momentsL}
    \overline{\mathcal{F}}_L(\omega,Q^2) &= \sum_{n=1}^\infty 2\omega^{2n}M^{(L)}_{2n}(Q^2), &&\text{with} \; M^{(L)}_{2n}(Q^2)= \int_{0}^1 dx\,x^{2n-2} F_L(x,Q^2).
\end{align}

\section{Feynman-Hellmann technique}
Our implementation of the second order Feynman-Hellmann method is presented in detail in~\cite{PhysRevD.102.114505}. Here, we briefly summarise its main aspects. 

An analysis of the Compton amplitude, such as the one given in \Cref{eq:compamp}, requires the evaluation of lattice 4-point functions. Application of the Feynman-Hellmann method reduces this problem to a more simple analysis of 2-point correlation functions using the established techniques of spectroscopy. To start, we modify the fermion action with the following perturbing term,
\begin{equation}\label{eq:fh_perturb}
    S(\lambda) = S + \lambda \int d^3z (e^{i \bv{q} \cdot \bv{z}} + e^{-i \bv{q} \cdot \bv{z}}) \mathcal{J}_{\mu}(z) ,
\end{equation}
where $\lambda$ is the strength of the coupling between the quarks and the external field, $\mathcal{J}_{\mu}(z) = Z_V \bar{q}(z) \gamma_\mu q(z)$ is the electromagnetic current coupling to the quarks, $\bv{q}$ is the external momentum inserted by the current and $Z_V$ is the renormalization constant for the local electromagnetic current. 

The main strategy to derive the relation between the energy shift and the matrix element is to work out the second-order derivatives of the two-point correlation function with respect to the external field from two complementary perspectives. Differentiating the energy of the perturbed nucleon correlator, $G^{(2)}_\lambda(\bv{p};t) \simeq A_\lambda(\bv{p}) e^{-E_{N_\lambda}(\bv{p}) t}$,  one finds a distinct temporal signature for the second-order energy shift, and by matching it to the expression, coming from a direct evaluation of the correlator, one arrives at the desired relation between the energy shift and the matrix element describing the Compton amplitude,
\begin{equation} \label{eq:secondorder_fh}
    \left. \frac{\partial^2 E_{N_\lambda}(\bv{p})}{\partial \lambda^2} \right|_{\lambda=0} = - \frac{T_{\mu\mu}(p,q) + T_{\mu\mu}(p,-q)}{2 E_{N}(\bv{p})},
\end{equation}
where $T$ is the Compton amplitude defined in \Cref{eq:compamp}, $q=(0,\bv{q})$ is the external momentum encoded by \Cref{eq:fh_perturb}, and $E_{N_\lambda}(\bv{p})$ is the nucleon energy at momentum $\bv{p}$ in the presence of a background field of strength $\lambda$. This expression is the principal relation that we use to access the Compton amplitude and hence the Compton structure functions as in \Cref{eq:compF1,eq:compF2}. For a more detailed derivation, see~\cite{PhysRevD.102.114505}.

\section{Selected results and discussion}
We carry out our simulations on QCDSF/UKQCD-generated $2+1$-flavour gauge configurations. Two ensembles are used with volumes $V=[32^3 \times 64, 48^3 \times 96]$, and couplings $\beta=[5.50, 5.65]$ corresponding to lattice spacings $a=[0.074, 0.068] \, {\rm fm}$ and the physical cut-offs $a^{-1} = [2.667, 2.902] \, {\rm GeV}$, respectively. Quark masses are tuned to the $SU(3)$ symmetric point where the masses of all three quark flavours are set to approximately the physical flavour-singlet mass, $\overline{m} = (2 m_s + m_l)/3$~\cite{Bietenholz:2010jr,Bietenholz:2011qq}, yielding $m_\pi \approx [470, 420] {\rm MeV}$. We obtain amplitudes for several values of current momentum, $Q^2$, in the range $1.5 \lesssim Q^2 \lesssim 7$ GeV$^2$. Multiple $\omega$ values are accessed at each simulated value of $\bv{q}$ by varying the nucleon momentum $\bv{p}$, which allows for a mapping of the $\omega$ dependence of the Compton structure functions. 

In order to extract the second order energy shift from the lattice correlation functions, we construct the following ratio,
\begin{equation} 
    \label{eq:ratio}
    \mathcal{R}^e_\lambda (\bv{p},t) \equiv \frac{G^{(2)}_{+\lambda}(\bv{p},t) G^{(2)}_{-\lambda}(\bv{p},t)}{\left( G^{(2)}(\bv{p},t) \right)^2}
    \xrightarrow{t \gg 0} A_\lambda(\bv{p}) e^{-2\Delta E^e_{N_\lambda}(\bv{p}) t},
\end{equation}
which isolates the energy shift only at even orders of $\lambda$, $\Delta E_{N_\lambda}^e(\bv{p})$, where $G^{(2)}_{\pm\lambda}(\bv{p},t)$ are the perturbed two-point functions and $G^{(2)}(\bv{p},t)$ is the unperturbed one. We compute the perturbed two-point correlation functions with two values of $|\lambda| \leq 0.025$.

\begin{figure}[t]
    \centering
    \includegraphics[width=.495\textwidth]{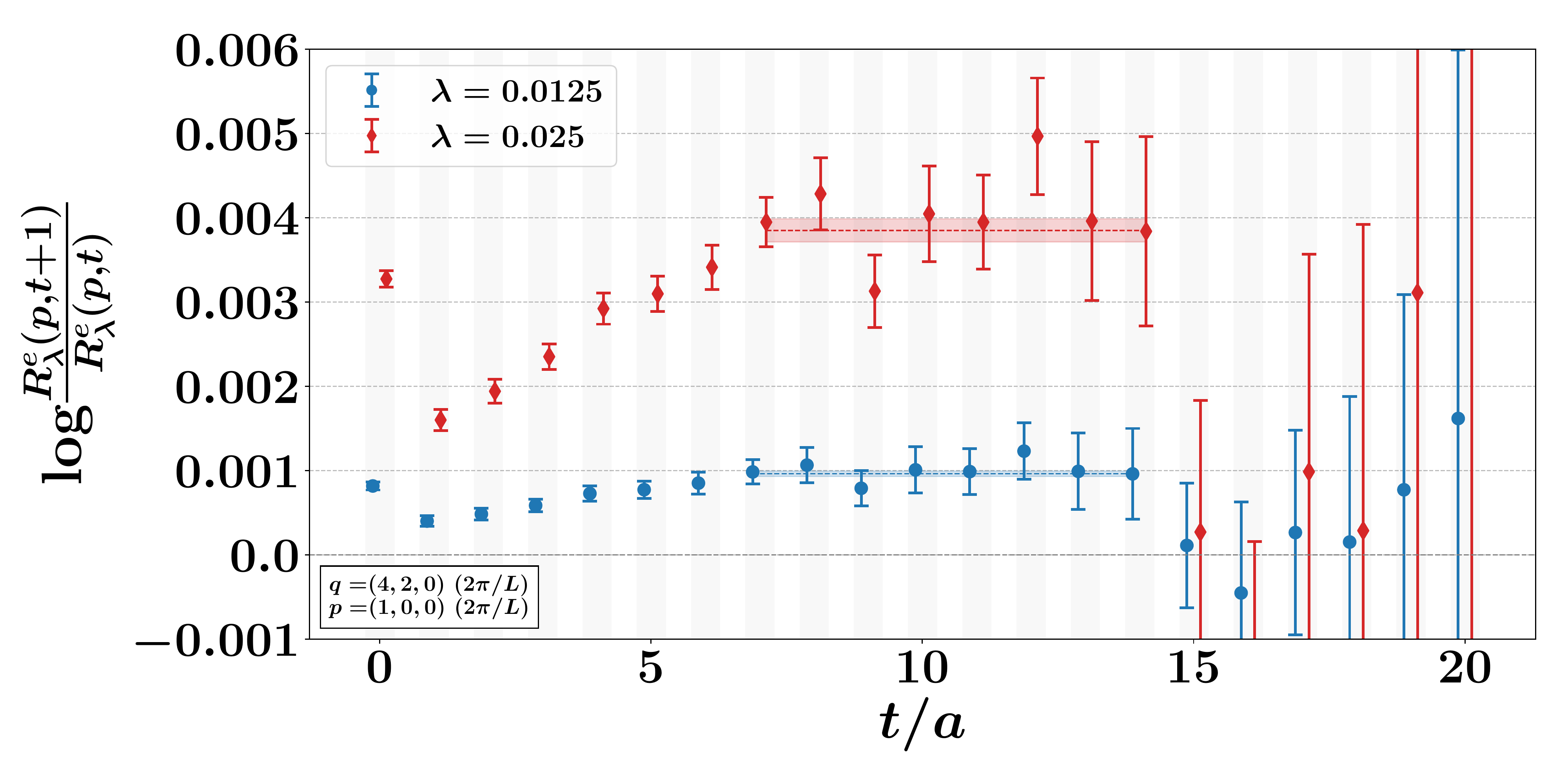}
    \includegraphics[width=.495\textwidth]{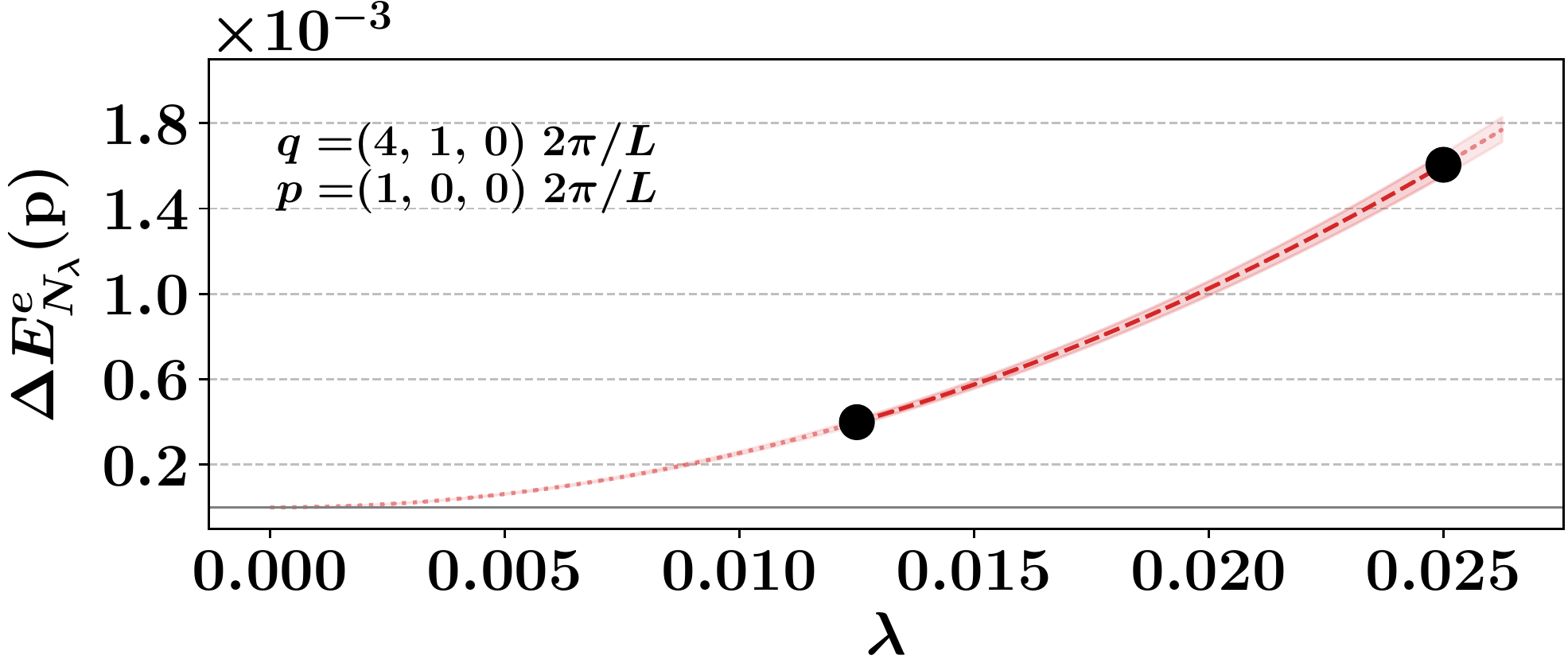}
    \caption{\label{fig:eshift}Left: Effective mass plot of the ratio given in \Cref{eq:ratio} for a $(\bv{q},\bv{p})$ pair. Data points are shifted for clarity. Right: $\lambda$ dependence of $\Delta E^e_{N_\lambda}(\bv{p})$. Error bars are smaller than the symbols. Plots taken from \cite{PhysRevD.102.114505}.}
\end{figure}

Having even-$\lambda$ energy shifts at two $\lambda$ values, we perform polynomial fits of the form, $\Delta E^e_{N_\lambda}(\bv{p}) = \frac{\lambda^2}{2} \left. \frac{\partial^2 E_{N_\lambda}(\bv{p})}{\partial \lambda^2} \right|_{\lambda = \bv{0}} + \mathcal{O}(\lambda^4)$, 
to determine the second order energy shift. The unperturbed energy, $E_N$, and odd-order lambda terms ($\mathcal{O}(\lambda)$, $\mathcal{O}(\lambda^3)$, $\dots$) are removed by construction in the ratio~(\ref{eq:ratio}). Given the smallness of our $\lambda$ values, higher order $\mathcal{O}(\lambda^4)$ terms are heavily suppressed, hence the fit form reduces to a simple one parameter polynomial.  We show representative cases for the signal quality and the $\lambda$ fits in \Cref{fig:eshift} from the $32^3 \times 64$ ensemble. 

The above analysis is performed to map out the $\omega$ dependence of the Compton structure functions given in \Cref{eq:compF1,eq:compF2} for each $Q^2$ value that we study. $\mathcal{F}_L(\omega,Q^2)$ is constructed according to \Cref{eq:FL_comp}. Once we extract the data points, we perform a simultaneous fit of $\overline{\mathcal{F}}_1$ and $\mathcal{F}_2$ in a Bayesian framework to determine the first few Mellin moments of the structure functions, Eqs.~(\ref{eq:ope_moments1}) to~(\ref{eq:ope_momentsL}), where we truncate the series at $n=4$. No dependence on higher-order terms is seen. Note that $\mathcal{F}_2$ is parametrised in terms of $\overline{\mathcal{F}}_1$ and $\mathcal{F}_L$ following \Cref{eq:FL_comp} since the Compton structure functions $\mathcal{F}_{1,L}$ are directly related to the cross sections and we can impose the positive-definiteness on the moments. We sample the moments from uniform distributions with bounds $M_{2}(Q^2) \in [0,1]$ and $M_{2n}(Q^2) \in [0, M_{2n-2}(Q^2)]$, for $n > 1$, to enforce the monotonic decreasing nature of the moments, $M_2(Q^2) \ge M_4(Q^2) \ge \cdot \cdot \cdot \ge M_{2n}(Q^2) \ge \cdot \cdot \cdot \ge 0$, for $u$ and $d$ contributions separately. The sequences of individual $uu$ or $dd$ moments are selected according to a multivariate probability distribution, $\operatorname{exp}(-\chi^2/2)$, where
$\chi^2 = \sum_{\mathcal{F}} \sum_{i,j} \left[ \mathcal{F}^\text{model}_i - \mathcal{F}^\text{obs}(\omega_i) \right] C^{-1}_{ij} \left[ \mathcal{F}^\text{model}_j - \mathcal{F}^\text{obs}(\omega_j) \right]$ 
is the $\chi^2$ function with the covariance matrix $C_{ij}$, ensuring the correlations between the data points are taken into account. We do not sample the isovector $uu-dd$ but instead construct it from the $uu$ and $dd$ pieces. Here, $\mathcal{F}$ stands for $\overline{\mathcal{F}}_1$ and $\mathcal{F}_2$, and the indices $i$, $j$ run through all the $\omega$ values and both flavours. 

\begin{figure}[t]
    \centering
    \includegraphics[width=\textwidth]{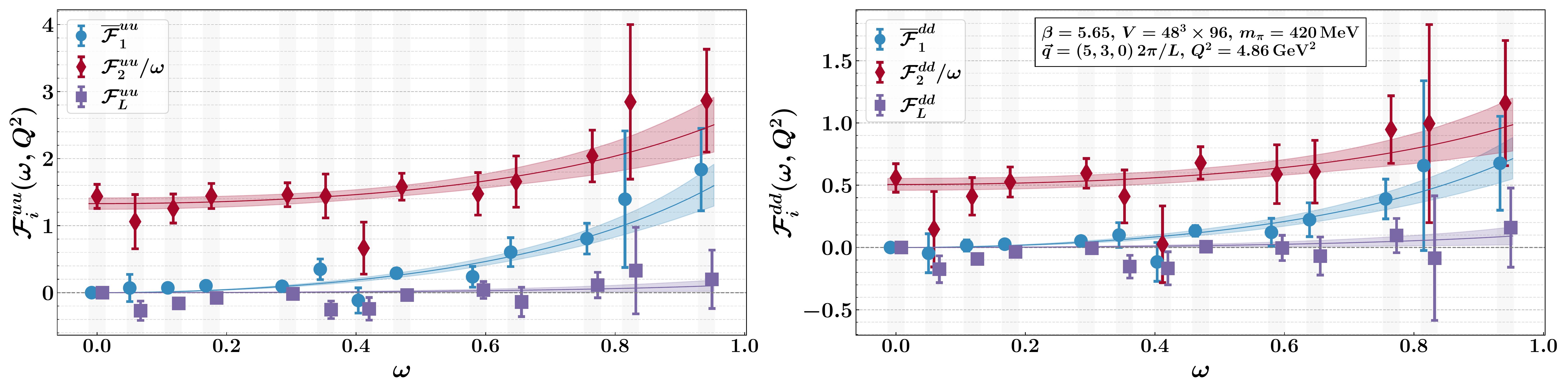}
    \caption{\label{fig:F12L_moments_highQ2}Fits to the $uu$ and $dd$ components of the Compton structure functions for $Q^2 = 4.86 \, {\rm GeV}^2$.}
\end{figure}
\begin{figure}[t]
    \centering
    \includegraphics[width=\textwidth]{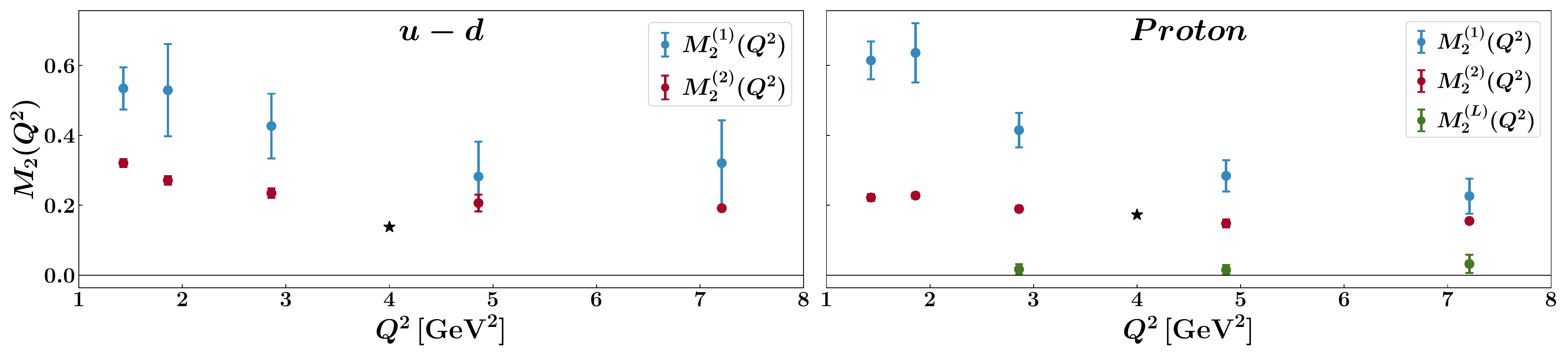}
    \caption{\label{fig:F12L_moments_Q2dep}$Q^2$ dependence of the first moments of $F_{1,2,L}$. We show the moments for the isovector $u-d$ and proton structure functions. Black stars are the experimental Nachtmann moments of $F_2$ at $Q^2 = 4 {\rm \, GeV}^2$ taken from~\cite{E06-009:2018mzv}.}
\end{figure}
Compton structure functions and fits depicting the extraction of the moments are shown in \Cref{fig:F12L_moments_highQ2} for a representative case from the $48^3 \times 96$ ensemble. 
We show the isovector $u-d$ moments of $F_{1}$ and $F_{2}$, and the moments for the proton $F_{1}$, $F_{2}$, and $F_{L}$ in \Cref{fig:F12L_moments_Q2dep}. Contributions from the $uu$, $dd$ and $ud$ (not shown) pieces are weighted with their electric charges and combined in obtaining the proton results. 

At low- and mid-$Q^2$ values, i.e. $Q^2 \lesssim 4 {\rm \, GeV}^2$, moments of $F_1$ and $F_2$ are significantly different from each other. As we approach the larger $Q^2$ values, i.e. $Q^2 > 4 {\rm \, GeV}^2$, they tend to similar asymptotic values close to their respective phenomenological values. Additionally, closing of the gap between the $F_1$ and $F_2$ moments can be interpreted as the recovery of the Callan-Gross relation, which is supported by the almost vanishing moments of $F_L$.

\section{Conclusions}
We have presented QCDSF/UKQCD Collaboration's recent efforts on calculating the Compton amplitude directly on the lattice via a novel extension of the Feynman-Hellmann techniques. Accessing the Compton amplitude allows us to extract the moments of the nucleon structure functions at a range of photon virtualities. We have shown preliminary results of the moments of the unpolarised $F_1$, $F_2$ and $F_L$ structure functions along with their $Q^2$ dependence. These results, especially for $F_L$, are a first for lattice structure function calculations, where we are able to extract the unpolarised structure functions simultaneously and study their behaviour at low- and mid-$Q^2$ values. 

\acknowledgments
The numerical configuration generation (using the BQCD lattice QCD program~\cite{Haar:2017ubh})) and data analysis (using the Chroma software library~\cite{Edwards:2004sx}) was carried out on the DiRAC Blue Gene Q and Extreme Scaling (EPCC, Edinburgh, UK) and Data Intensive (Cambridge, UK) services, the GCS supercomputers JUQUEEN and JUWELS (NIC, Jülich, Germany) and resources provided by HLRN (The North-German Supercomputer Alliance), the NCI National Facility in Canberra, Australia (supported by the Australian Commonwealth Government) and the Phoenix HPC service (University of Adelaide). RH is supported by STFC through grant ST/P000630/1. HP is supported by DFG Grant No. PE 2792/2-1. PELR is supported in part by the STFC under contract ST/G00062X/1. GS is supported by DFG Grant No. SCHI 179/8-1. KUC, RDY and JMZ are supported by the Australian Research Council grant DP190100297.

\providecommand{\href}[2]{#2}\begingroup
\renewcommand{\baselinestretch}{1}
\setlength{\bibsep}{.25pt}
\setstretch{1}
\footnotesize
\raggedright
\endgroup

\end{document}